\documentclass[aps,prl,twocolumn]{revtex4-1}
\usepackage{amssymb}
\usepackage{amsmath,bm}
\usepackage{graphicx}
\usepackage[usenames,dvipsnames]{color}
\usepackage{bbm}
\usepackage{cancel}
\usepackage{ulem}
\usepackage{multirow}
\newcommand{\B}[1]{{\bm{#1}}}

\def\d{{\rm d}}

\begin{document}
\normalem

\title{Frictional Granular Matter: Protocol Dependence of Mechanical Properties}

\author{Ana\"el Lema\^{\i}tre$^1$,  Chandana Mondal$^2$, Itamar Procaccia$^{2,3}$ Saikat Roy$^2$, Yinqiao Wang $^4$ and Jie Zhang $^{4,5}$}
\affiliation{$^1$NAVIER, UMR 8205, \'Ecole des Ponts ParisTech, IFSTTAR, CNRS, UPE, Champs-sur-Marne, France. \\ $^2$Department of Chemical Physics, the Weizmann Institute of Science, Rehovot 76100, Israel. \\ $^3$Center for OPTical IMagery Analysis and Learning, Northwestern Polytechnical University, Xi'an, 710072 China.\\ $^4$School of Physics and Astronomy, Shanghai Jiao Tong University, 800 Dong Chuan Road, 200240 Shanghai, China.\\$^5$ Institute of Natural Sciences, Shanghai Jiao Tong University, 200240 Shanghai, China.}

\begin{abstract}
Theoretical treatments of frictional granular matter often assume that it is legitimate to invoke classical elastic theory
to describe its coarse-grained mechanical properties.
Here we show, based on experiments and numerical simulations, that this is generically not the case since stress auto-correlation functions decay more slowly than the elastic Green's function. It was shown theoretically that standard elastic decay demands pressure and torque density fluctuations to be normal, with possibly one of them being hyperuniform.
Generic compressed frictional assemblies exhibit however abnormal pressure fluctuations, failing to conform with the central limit theorem.
The physics of this failure is linked to correlations built in the material during compression from a dilute configuration prior to jamming.
By changing the protocol of compression one can observe different pressure fluctuations and stress auto-correlations decay at large scales.
\end{abstract}

\maketitle

{\bf Introduction}:
Frictional forces are usually at work in assemblies of macroscopic particles, known as granular materials, widely present in the energy, pharmaceutical, chemical and food industries, as well as in the environment.  Understanding the properties of these frictional granular solids is of fundamental and practical importance. Traditionally, engineers have used elasticity theory \cite{05Ned,98Sav} to describe granular bulk properties at low loads, and elasto-plastic models \cite{05Ned,98Sav,83JM,02GG} to describe yield and quasi-static flow. In this Letter we show that \emph{even at mechanical equilibrium} the predictions of elasticity theory may be at variance with the realities of frictional granular matter. We present both experiments and numerical simulations (which are in agreement with each other) to substantiate this claim.
\begin{figure}
	\includegraphics[scale=0.40]{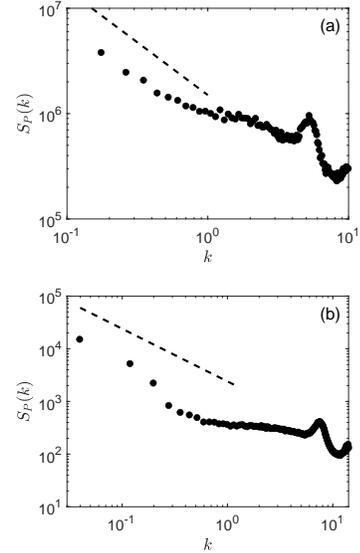}
	\caption{The failure of elasticity theory in frictional granular matter. Shown are plots of the pressure autocorrelation function $S_{P}(\B{k})$ vs. k. Panels a and b display experimental and simulational results respectively, see text for details.  The dashed line in both panels represents the power law $k^{-1}$.}
	\label{spectrum}
\end{figure}

Fundamental to elasticity theory are the properties of the stress tensor, especially the correlation function of its components and their fluctuations.
Consider a two-dimensional assembly of $N$ granules in an area $A$ whose centers of masses are at positions $\B r_i$. The $j$th granule exerts on every neighboring $i$th granule a force $\B F_{ij}$ which may have both normal $\B F^{(n)}_{ij}$ and tangential (frictional) component $\B F^{(t)}_{ij}$.
The Cauchy stress tensor, which is by definition the main contribution to the momentum flux~\cite{90ECM,02GG} reads:
\begin{equation}\label{eq:gg}
{\sigma}_{\alpha\beta}(\B r)=-\frac{1}{2}\sum_{i,j; i\ne j} F_{ij}^\alpha r_{ij}^\beta\int_0^1\d s\,\phi(\B r-\B r_i+s\B r_{ij})
\end{equation}
where $\alpha,\beta$ refer to Cartesian coordinates and $\B r_{ij}=\B r_i-\B r_j$ is the difference vector between disk centers. Here, $\phi$ is a spatially isotropic function of unit integral: it may either be taken as the 2D delta function $\delta^{(2)}$~\cite{90ECM} (one then speaks of Hardy's stress~\cite{90ECM}) or as a finite-range function to introduce coarse-graining~\cite{02GG}. In Fourier space, the same expression reads:
\begin{equation}\label{eq:hardy}
\widehat{\sigma}_{\alpha\beta {\B k}}=\frac{1}{2A}\,\widehat{\phi}(\B k)\sum_{i,j,i\ne j}\,F_{ij}^{\alpha}r_{ij}^{\beta}\frac{e^{-i{\B k}\cdot {{\B r}_i}}-e^{-i{\B k}\cdot {{\B r}_j}}}{i\B k\cdot {\B r}_{ij}}
\end{equation}
which evidences that the limit $\phi\to\delta^{(2)}$, i.e. $\widehat\phi\to1$, is well-behaved and yields Hardy's stress.

  When elasticity prevails, the pressure autocorrelation $S_{P}(k)$ (with pressure $P\equiv-\frac{1}{2}({\sigma}_{xx}+{\sigma}_{yy})$) should reach a constant value when $k\to 0$~\cite{17Lem,18Lem}.
  In Fig.~\ref{spectrum} we show both experimental and simulational results for compressed frictional
granular matter in mechanical equilibrium. The interesting result is that $S_{P}(\B{k})$ actually displays a divergence in the limit $k\to 0$ which appears as a scaling law
\begin{equation}
S_{P}(k) \sim k^{-\nu} \ , \quad \nu\approx 1 \ .
\label{diverge}
\end{equation}
Thus these frictional assemblies are in disagreement with elasticity theory. We turn now to describe the experiment and simulations that
resulted in Fig.~\ref{spectrum}.

{\bf Materials and Methods}: The experiment employs a biaxial apparatus to prepare isotropically jammed packings of photoelastic disks (in fact, flat cylinders), allowing us to measure all the  individual vector contact forces. The apparatus consists of a square frame mounted on top of a powder-lubricated horizontal glass plate.  The frame is filled with a 50:50 mixture of 5000 bi-dispersed photoelastic disks (Vishay PSM-4), with diameters of 1.4 cm and 1.0 cm. Below the experimental results are displayed using the diameter of the smaller particle as the unit of length. Each pair of parallel walls of the square frame can move symmetrically with a motion precision of 0.1 mm such that the center of mass of the frame remains fixed. To apply isotropic compression, the two pairs of walls are programmed to move inwards symmetrically. The motion of walls, with a speed of 0.1$\mathrm{mm\ s^{-1}}$, is sufficiently slow to guarantee that the deformation process is quasi static. About $1.5$m above the apparatus, there is an array of 2$\times$2 high-resolution (100 pixel/cm) cameras that are aligned and synchronized. These four cameras record two different images. Disk positions are obtained from the {\em normal image}, recorded in the absence of a right-handed circular polarizer sheet placed below the cameras. Contact forces are analyzed from the {\em force-chain image}, recorded in the presence of a polarizer sheet below the cameras, using a force-inversion algorithm \cite{05MB}.  Right below the glass plate of the apparatus, a left-handed polarizer sheet is attached, below which a light panel is placed to provide a uniform illumination of the disk packings within the square frame. The experimental result shown in Fig.~\ref{spectrum} was obtained
starting from a random dilute configuration of disks and applying isotropic compression till a target pressure $P=20\ \mathrm{N\ m^{-1}}$ is reached. The corresponding packing fraction is around $82.3 \%$. In these conditions the friction of our disks with the lubricated substrate is about 36 time smaller
than the typical contact force. 

The simulations use amorphous granular assemblies of $16000$ disks, half of which have a radius $R_1=0.35$ and the other half
with a radius $R_2=0.49$ (with the same ratio of 1.4 as in the experiment). The contact forces, which include both normal and tangential components due to friction, are modeled according to the discrete element method developed by Cundall and Strack \cite{79CS}, combining a Hertzian normal force and a tangential Mindlin component. Full details of these forces and the equations of motion solved can be found in Refs.~\cite{01SEGHLP,19CGPP,19CGPPa,20LMPR}. Simulations are performed using the open source codes, LAMMPS \cite{95Pli} and LIGGGHTS \cite{12KGHAP} to properly keep track of both the normal and the history-dependent tangential force. Initially, the grains are placed randomly in a large two dimensional box while forbidding the existence of overlaps or contacts. The system is then isotropically compressed along $x$ and $y$ directions while integrating Newton's second law with total forces and (scalar) torques acting on particle $i$ given by
$\B F_{i}= \sum_{j}\B F^{(n)}_{ij} + \B F^{(t)}_{ij}$, and
$\tau_{i}= \sum_{j}\tau_{ij}$
with
\begin{equation}
\tau_{ij}\equiv-\frac{1}{2}\left({\B r}_{ij} \times\B F^{(t)}_{ij}\right)\cdot{\B e}_z
\end{equation}
the torque exerted by $j$ onto $i$.
Compression is performed using a series of steps which involve: (i) one MD step during which we reduce the box lengths along $x$ and $y$ directions by 0.002\%; (ii) a constant NVE run, until the force and torque on each and every particle are smaller than $10^{-7}$ in reduced units. This guarantees that the cell remains square throughout the process. We repeat these compression and relaxation cycles until the system attains a jammed (mechanically balanced) configuration at the chosen final pressure, fixed to 72.0 (in reduced units)~\cite{20LMPR}.
Of course, in the final \emph{mechanically equilibrated states} obtained at the end of compression the total forces and torques $\B F_i$ and $\tau_i$ vanish with $10^{-7}$ accuracy, as well as all the velocities.

{\bf Theoretical background and consequence}:
\emph{In the absence of friction}, since $\B F_{ij}$ and $\B r_{ij}$ are colinear for any pair $(i,j)$, stress is symmetric. It was then shown that the two conditions of (i)~mechanical balance and (ii)~material isotropy imply that the full tensorial stress autocorrelation is completely determined by the pressure autocorrelation only~\cite{17Lem,18Lem}. Moreover, due to the nature of this relation, the elastic ($1/r^2$) decay of stress correlation then follows from the normality of pressure fluctuations. This condition is defined by considering the average pressure on circles of radius $R$, and computing the variance $V_P(R)$ due to circle-to-circle and sample-to-sample fluctuations:
\begin{equation}
V_P(R) \equiv \langle P(R)^2 \rangle - \langle P(R)\rangle ^2 \sim \frac{1}{R^\eta}\ .
\label{eqvar}
\end{equation}
When the pressure has normal fluctuations this variance is expected to decay like the inverse area of the averaging domain, i.e. as $1/R^2$. When it does,
\begin{equation}
\lim_{k\to 0} S_{P}(\B{k}) = Const \ ,
\label{normdec}
\end{equation}
without any divergence \cite{17Lem,18Lem}.
More generally, the exponent $\nu$ of Eq.~(\ref{diverge}) satisfies $\nu=2-\eta$ \cite{20LMPR}.

\begin{figure}
	\includegraphics[scale=0.40]{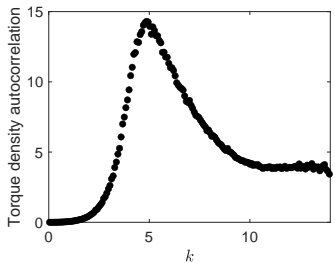}
	\caption{Torque density autocorrelation function as a function of $k$ as compute in the numerical simulation. The fact that this autocorrelation
		function {\em vanishes} in the limit $k\to 0$
		 shows that the torque density is hyperuniform.}
	\label{torque}
\end{figure}

\emph{In frictional granular assemblies}, the local torque density field $\frac{1}{2}({\sigma}_{xy}-{\sigma}_{yx})$ is not identically zero due to the existence of contact torques---stress is no longer symmetric. It remains that conditions~(i) and~(ii) strongly constrain the full stress autocorrelation, yet not to the same extent: it is now fully determined by the autocorrelations of \emph{both} pressure and torque density, which are two spatially isotropic functions~\cite{20LMPR}. Moreover, normal (elastic, $\propto1/r^2$) correlation decay now demands that the fluctuations of pressure and torque density are normal, with one of them possibly hyperuniform. We present in Fig.~\ref{torque} the torque density autocorrelation function computed in the same ensemble leading to Fig.~\ref{spectrum}  lower panel. Obviously, the torque density is hyperuniform \cite{20LMPR}. It follows that the long-range stress correlation decay is fully determined by pressure only. Hence the data of Fig.~\ref{spectrum} demonstrate that stress correlations are inconsistent with elastic behavior.

\begin{figure}
	\includegraphics[scale=0.35]{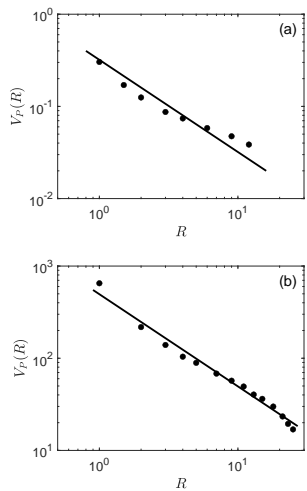}
	\caption{Pressure variance as a function of sampling radius. Upper panel: experiment. Lower panel: simulation. In both panels
	the straight line represents the power law $R^{-1}$}
	\label{variance}
	\end{figure}

{\bf Pressure fluctuations}: The pressure variance was measured in both the simulations and the experiment as a function of $R$, cf. Fig.~\ref{variance}.
Clearly, in both cases the slope for large $R$ is much smaller than 2, and in rough agreement with $\eta\approx 1$ throughout
the range. This is perfectly consistent with the $1/k^\nu$ scaling of $S_P(k)$ in the small $k$ limit.

\begin{figure}
	\includegraphics[scale=0.38]{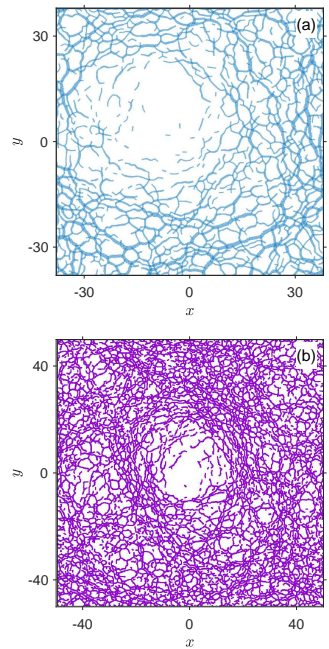}
	\caption{Force chains in compressed samples. Panel a: the experimental case, $N=5000$, $d$ is the average disk diameter. Panel b:
		Simulations, $N=16000$. The packing fraction here is 0.819.}
	\label{forcechains}
\end{figure}

To visualize the reason for the failure of Central Limit Theorem we present the intense force chains in our frictional media. To this aim we compute the average magnitude of the forces
$F_{ij}$, which is denoted as $\langle F_{ij} \rangle$, and then plot all the forces whose magnitude exceeds this
average (i.e $F_{ij} \ge \langle F_{ij} \rangle$). Two typical real space maps of these force chains, one from the
experiment and the other from the numerical simulations are shown in Fig.~\ref{forcechains}.
The point to notice is the glaring inhomogeneity which translates to anomalous correlation
functions as observed. It should be noted that periodic boundary conditions are used in the simulations: it emphasizes that the strong heterogeneity observed is not a consequence of domain boundaries, but an intrinsic feature of compressed frictional systems.

\begin{figure}
	\includegraphics[scale=0.40]{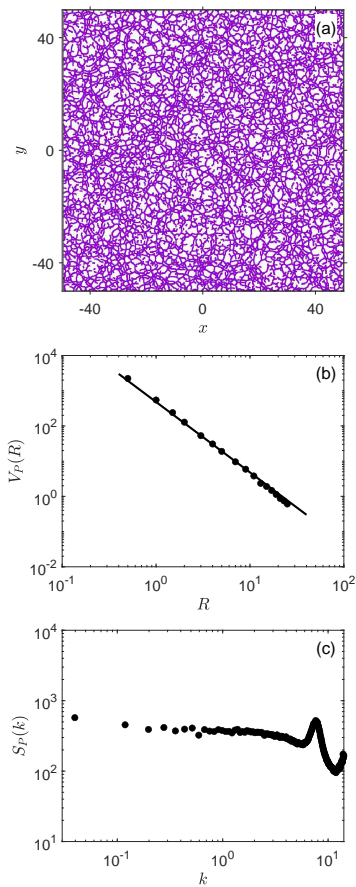}
	\caption{Results of simulations with switched off friction until jamming. Panel a:  Force chains in the compressed frictional assembly. The packing fraction here is 0.846.  Panel b: The pressure variance; here $\eta\approx 2$. Panel c: the pressure autocorrelation function. $N=16000$.}
	\label{switchoff}
\end{figure}

{\bf The creation of pressure inhomogeneity and correlations}: In light of the results shown above it becomes quite interesting to
identify at which stage of compression do we build up correlations that destroy the normality of pressure fluctuations. We recall that
our compression consists of two regimes, the first is before jamming and the second after. The numerical simulations provide us with
a very easy protocol to answer the question which regime is to be blamed. In the numerics we can switch off the friction between the
disks throughout the compression up to the jamming point, and then switch the friction on again for the final compression to the
target pressure. The results of this exercise are quite enlightening, cf. Fig.~\ref{switchoff}. In the three panels we show (a) the
force chains, (b) the pressure variance and (c) the pressure auto-correlation function.
We find that all the anomalies have disappeared. The force chains are homogeneous, the pressure fluctuation normal (here $\eta\approx 2$)
and the pressure auto-correlation function has a finite limit when $k\to 0$. \emph{All the anomalies are caused by the compression of the dilute
frictional disks before jamming!} Once jammed the correlations that were built up are frozen in and cannot relax.

\begin{figure}
	\includegraphics[scale=0.40]{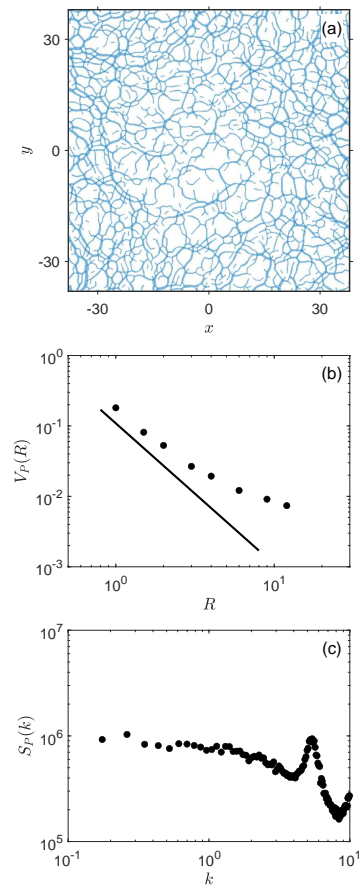}
	\caption{Results of experiments with mechanical agitation until jamming.  Panel a:  Force chains in the compressed frictional assembly. Panel b: The pressure variance; here the straight line corresponds to $\eta=2$. Panel c: the pressure autocorrelation function. $N=5000$.}
	\label{agitate}
\end{figure}

In the experiment we cannot switch off the friction in the dilute regime, but can mimic the effect of this protocol by trying to destroy
by hand the correlations built during this interval of compression. To do that, we start from a random dilute configuration of disks and apply isotropic compression quasi-statically in multiple steps of incremental strain $0.05\%$. After every 10 steps we apply small random mechanical agitation to the disk packing to eliminate (as much as possible) the force chains formed from the previous compression steps. At jamming the force chains that were created during the most recent compression step, though extremely weak, persist under the application of small mechanical agitation. This jamming packing fraction is typically close to the isotropic jamming point of packing fraction of $84\%$ which is typical to frictionless disks. We then apply quasi-static isotropic compression without mechanical agitation till a target pressure $P=20\ \mathrm{N\ m^{-1}}$. The corresponding packing fraction is around $85.3 \%$. Results analogous to those shown in Fig.~\ref{switchoff} are shown in Fig.~\ref{agitate}.
We conclude that while mechanical agitation is not as efficient as switching off the friction altogether, it does succeed to reduce considerably
the anomalies. The force chains appear more homogeneous, the pressure variance decays faster than before, and the pressure autocorrelations function
loses its divergence almost entirely.

{\bf Summary and discussion}: The most striking result of both the experiment and the simulations was that the pressure auto-correlations functions in the compressed frictional granular matter may exhibit a protocol-dependent divergence in the $k\to 0$ limit. Moreover, when divergences are present, the variance of the domain averaged pressure decays anomalously slowly.
Since elastic-like behaviour requires these functions, like the elastic Green's function, to converge to a constant at low $k$, and stress fluctuations to be normal,
our findings imply the surprising conclusion that a frictional granular assembly can exhibit elastic-like behavior or not depending on their preparation protocol.

Moreover, our experimental and numerical results demonstrate that the break-down of elasticity results from the build-up of some kind of structural correlations during the compression protocol, much before reaching jamming. Indeed, the protocol of compression started in both experiments and simulations with a dilute system of zero pressure, that was compressed isotropically until the system jammed, and then further compression brought it to a target pressure. We have thus discovered that the \emph{anomalous correlations form in the dilute phase, while the pressure was still zero}. Once the system jammed these anomalous correlations were already imprinted in the material and could not
be released. The signature is seen in the force chains that remain inhomogeneous while compressing after jamming.

{ \bf Acknowledgments}:
IP acknowledges the support of the US-Israel Binational Science Foundation and the scientific and cooperation agreement between Italy and Israel through the project COMPAMP/DISORDER. YW and JZ acknowledge support of the National Natural Science Foundation of China under grants No. 11774221 and 11974238.

\bibliography{All}

\end{document}